\documentclass[12pt,amsmath,amssymb,showpacs]{revtex4-1}
\usepackage{amssymb}
\usepackage{graphicx}
\usepackage{caption}
\usepackage{wrapfig}	
\usepackage{lipsum} 
\usepackage{varioref}
\usepackage{hyperref}
\usepackage{cleveref}
\usepackage{textcomp}
\usepackage{float}
\usepackage{epstopdf}
\usepackage{array}
\usepackage{csquotes}
\usepackage{xcolor}
\usepackage{diagbox}
\usepackage{makecell}
\usepackage{soul}
\usepackage{setspace}
\usepackage{mwe}
\usepackage{adjustbox}
\usepackage{multirow}
\AppendGraphicsExtensions{.tif}

\begin{document}

\title[]{Thickness induced metal to insulator charge transport and unusual hydrogen response in granular palladium nanofilms}

\author{Dharmendra K. Singh}
\affiliation{School of Physics, Indian Institute of Science Education and Research Thiruvanthapuram, Vithura, Thiruvananthapuram-695551, Kerala, India}
\author{Praveen S. G.}
\affiliation{School of Physics, Indian Institute of Science Education and Research Thiruvanthapuram, Vithura, Thiruvananthapuram-695551, Kerala, India}
\author{Adithya Jayakumar}
\affiliation{School of Physics, Indian Institute of Science Education and Research Thiruvanthapuram, Vithura, Thiruvananthapuram-695551, Kerala, India}
\author{Suma M. N.}
\affiliation{LPSC Bangalore, ISRO, Benguluru-560008, India}
\author{Vinayak B. Kamble}
\affiliation{School of Physics, Indian Institute of Science Education and Research Thiruvanthapuram, Vithura, Thiruvananthapuram-695551, Kerala, India}
\author{J. Mitra}
\affiliation{School of Physics, Indian Institute of Science Education and Research Thiruvanthapuram, Vithura, Thiruvananthapuram-695551, Kerala, India}
\author{D. Jaiswal-Nagar}
\email{deepshikha@iisertvm.ac.in}
\affiliation{School of Physics, Indian Institute of Science Education and Research Thiruvanthapuram, Vithura, Thiruvananthapuram-695551, Kerala, India}

\vspace{10pt}
\date{today}

\begin{abstract}
This work reports a systematic study of the evolution of charge transport properties in granular ultra-thin films of palladium of thickness varying between 6 nm and 2 nm. While the films with thickness $>$ 4 nm exhibit metallic behaviour, that at 3 nm thickness undergoes a metal-insulator transition at 19.5 K. In contrast, the 2 nm thick film remained insulating at all temperatures, with transport following Mott’s variable range hopping. At room temperature, while the thicker films exhibit resistance decrease upon H$_2$ exposure, the insulating film showed an anomalous initial resistance increase before switching to a subsequent decrease. The nanostructure dependent transport and the ensuing $H_2$ response is modeled on a percolation model, which also explores the relevance of film thickness as a macroscopic control parameter to engineer the desired system response in granular metal films.
\end{abstract}

\date{\today}

\maketitle


\section{Introduction}

Charge transport in granular conductors is an intense area of research, wherein, electronic properties of granular conductors can be tuned at the nanoscale by varying grain size and inter-grain separation \cite{murray,gaponenko,mowbray,bimberg,stangl}. Grain size distribution in a granular conductor is known to vary from a few to hunderds of nanometers, resulting in characteristics originating from quantization of confined electron states at few nanometer scale and collective properties of coupled grains at hundreds of nanometer scale, thus, opening the possibility of a variety of novel and improved applications ranging from electronic to optoelectronic \cite{bimberg,stangl}. Ultra-thin films grown by physical vapour deposition, constitute an important method of preparation of such granular conductors \cite{barr,wu,morris}. It is expected that as the parameters of the deposition are changed such that a transformation from a continuous film to an island-like configuration results, electrical transport may correspondingly change due to quantum confinement effects arising due to an interplay of opposing effects of length scales and energy scales. Such an effect has, in fact, been observed in different systems ranging from self-assembled quantum dots \cite{yakimov,murray,gaponenko,mowbray,bimberg,stangl}, granular metal films \cite{kubo,gorkov,chuang,zabrodskii,khondaker}, nanocluster assembled films \cite{bansal,tejal,praveen} etc. Additionally, charge conduction in a disordered metal is known to vary significantly from that of a pure metal, wherein, mechanisms like inelastic electron scattering from impurities and defects contribute significantly to charge transport apart from the ``pure'' electron-phonon scattering observed in pure metals \cite{gershenzon,bergmann,pritsina}. So, it is expected that charge transport in granular metal films, where disorder in a given film changes as the thickness is varied, would also be affected by such mechanisms.\\       
Palladium (Pd) metal offers a possibility with respect to charge transport in granular metals, since Pd metal is, additionally, well known for catalytic activity, especially for hydrogen (H$_2$) gas and selectively absorb it  \cite{flanagan,kay,zuhner,bohmholdt,sakamoto,wolf,barr,wu,morris,favier,dankert,lee,ramanathan,zeng,walter,yang,jiang,cabrera,xu,krishnan,kumar,raviprakash,mitra,feng,dawson}, thus, providing an extra parameter that can be tuned to control the charge transport mechanism in nano-sized grains of Pd films. In bulk Pd and at room temperature, the incoming H$_2$ molecules are physisorbed on the Pd surface and dissociate into hydrogen (H) atoms due to the high reactivity of the Pd atoms to H atoms \cite{kay}, which then, diffuse into the Pd lattice until they reach the octahedral sites of face centred cubic (fcc) Pd \cite{zuhner,bohmholdt}. The process of diffusion is enhanced at the grain boundaries or dislocations since they provide a high diffusivity path \cite{sakamoto,lee}. The random occupation of H into the Pd lattice results in a solid solution of Pd and H called the $\alpha$-phase (PdH$_{\alpha}$) and extends for exposure of Pd till $\sim$ 15,000 ppm H$_2$ concentration \cite{narehood}. In the $\alpha$-phase, Pd lattice expands by 0.15$\%$ due to incorporation of H atoms in the interstitial sites of Pd. Above $\sim$ 15,000 ppm H$_2$ concentration, a phase transformation happens between $\alpha$ phase to $\beta$ phase (PdH$_{\beta}$) that results in a lattice constant increase from 3.895 \AA (maximum for $\alpha$ phase) to 4.025 \AA (minimum for $\beta$ phase), resulting in a 3.4$\%$ increase in the lattice size \cite{lewis,flanagan}. The PdH$_x$ structures above 15,000 ppm H$_2$ concentrations efficiently scatter conduction electrons leading to resistance increase of the material. In continuous Pd thin films grown by various methods like sputtering, thermal evaporation and pulsed laser deposition, an increase in resistance with  exposure has, in fact, been reported \cite{cabrera,krishnan,kumar,raviprakash}. It was also found by Lee et al. \cite{lee} that, in an ultra- thin film form, the process of $\alpha$ to $\beta$ structural transition is hysteretic, with the width of the hysteresis loop, strongly dependent on the film thickness. However, for ultra-thin films of 5 nm thickness, no hysteresis was found.\\
If the ultra-thin Pd films are made discontinuous with the size of each island in the nanometer range and the inter-island separation lesser than a nanometer, then charge transport may happen via tunneling \cite{praveen,coutts,simmons}. It was found that if such discontinuous films are near the percolation threshold, then H$_2$ exposure would have two effects on the charge transport: (i) H$_2$ adsorbed on the surface would result in an increase in the work function of the island, resulting in an increase in resistance of the film and (ii) expansion of the Pd islands leading to a decrease of the inter-island separation, thereby, decreasing the effective resistance of the assembly \cite{barr,wu,morris}. Using this process of hydrogen absorption-induced lattice expansion (HAILE), a decrease in resistance has been observed in a variety of Pd configurations ranging from thin films \cite{dankert,lee,xu,ramanathan,wu} to nanowires \cite{zeng,walter,yang}, nanofibres \cite{jiang}, nanoclusters \cite{shin} etc. However, fundamental understanding of the relationship between film thickness, grain size, resistivity and sensitivity of ultra-thin Pd films is far from complete. In this investigation, we have undertaken a systematic study of the electronic properties of ultra-thin Pd films with mass equivalent thickness in the range 2 nm to 6 nm, and their change after exposure to H$_2$ gas via measurement of the change in their resistive response. Before being exposed to H$_2$, temperature dependent resistance measurements showed the 6 nm, 5 nm and 4 nm thin films to be metallic (dR/dT $>$ 0) with the 3 nm thin film exhibiting a metal to insulator transition at $\sim$ 19.5 K. In contrast, the 2 nm films were found to be insulating and Mott’s variable range hopping mechanism was found to govern the charge transport in these films. Importantly, all the films that were metallic at room temperature (thickness 6 - 3 nm) were found to exhibit a decrease in resistance on being exposed to H$_2$ due to HAILE phenomenon. However, the 2 nm thick films that were insulating at room temperature were found to have an initial increase in resistance upon exposure to hydrogen, followed by a subsequent decrease. We found that the time constants needed to reach $\sim$ 83$\%$ of the initial resistance were two rather than one. In order to explain the existence of two time-constants in our ultra-thin Pd films, we propose a model employing percolation induced opening up of new conduction pathways.  

\section{Experimental Details}
Gold, Chromium and Palladium wires used in thermal evaporator were purchased from M/s. Goodfellow Cambridge Ltd., U.K. Pd films constituting nanoscale islands were deposited on a pre-fabricated structure as shown in Fig. \textbf{\ref{AFM} (a)} using a commercial electron beam evaporator (Tectra e-flux), attached to an ultra-high vacuum (UHV) chamber with turbo molecular pumps. The base pressure achieved in the chamber prior to the deposition was $\sim$ 3 x 10$^{-7}$ mbar. The films were deposited at the rate 0.2 A/s at a chamber pressure $\sim$ 2 x 10$^{-6}$ mbar. The thicknesses and deposition rate of the films were monitored using a quartz crystal microbalance (Inficon), in the evaporation chamber. The reported thicknesses are “nominal” as displayed values since a true thickness would be poorly defined for a quasi-continuous nano-island film. The pre-fabricated substrates used for the deposition of nano-island films and shown in Fig. \ref{AFM} (a), were prepared as follows: First, pre cleaned glass slides were cut in to 1 cm x 3 mm size and the centre of the glass slides was shadow masked using a thin copper wire of thickness 100 $\mu$m. These shadow masked glass slides were loaded in a thermal evaporator to deposit gold layers of 50 nm thickness, to be used as contact pads. Prior to deposition of the gold layers, a thin wetting layer of chromium was also deposited. After making the contact pads, the shadow mask was removed and a channel region of 80 $\mu$m width was obtained. Pd nano-island films were deposited on the channel regions of these structures. Morphological characterisation of the as deposited Pd films were conducted using a Bruker Multimode 8 Atomic Force Microscope (AFM) while the crystallinity of the grown films was investigated using 300 keV FEI Tecnai F30’s high resolution transmission electron microscope (HRTEM).\\
Resistance (R) vs. temperature (T) measurements were carried out on Nanomagnetics Hall effect measurement system’s closed cycle refrigerator. Before measuring R-T for a given film, it was ensured that the current-voltage (I-V) curves for a given film were linear in the $\pm$ 100 mV range both at 300 K as well as 2.1 K. The base temperature achievable in this system is 3.6 K with a resolution of $\pm$ 0.2 K. Change of transport mechanism in Pd ultra-thin films upon exposure to hydrogen gas was studied using an in-house built set-up \cite{suresh}. The H$_2$ gas used in the measurement set-up was air balanced. Both the H$_2$ gas flow as well as air flow towards the measurement chamber were regulated using Alicat Scientific’s mass flow controllers (MFC's) to achieve desired dilution of the H$_2$ gas with synthetic air. The change of resistance of the Pd films upon the sorption of H$_2$ was measured using Keithley’s 2700 Digital multimeter cum data acquisition system. 

\section{Results and Discussion:}
\subsection{HRTEM measurements}
\begin{figure}[h!]
	\centering
		\includegraphics[width=0.94\textwidth]{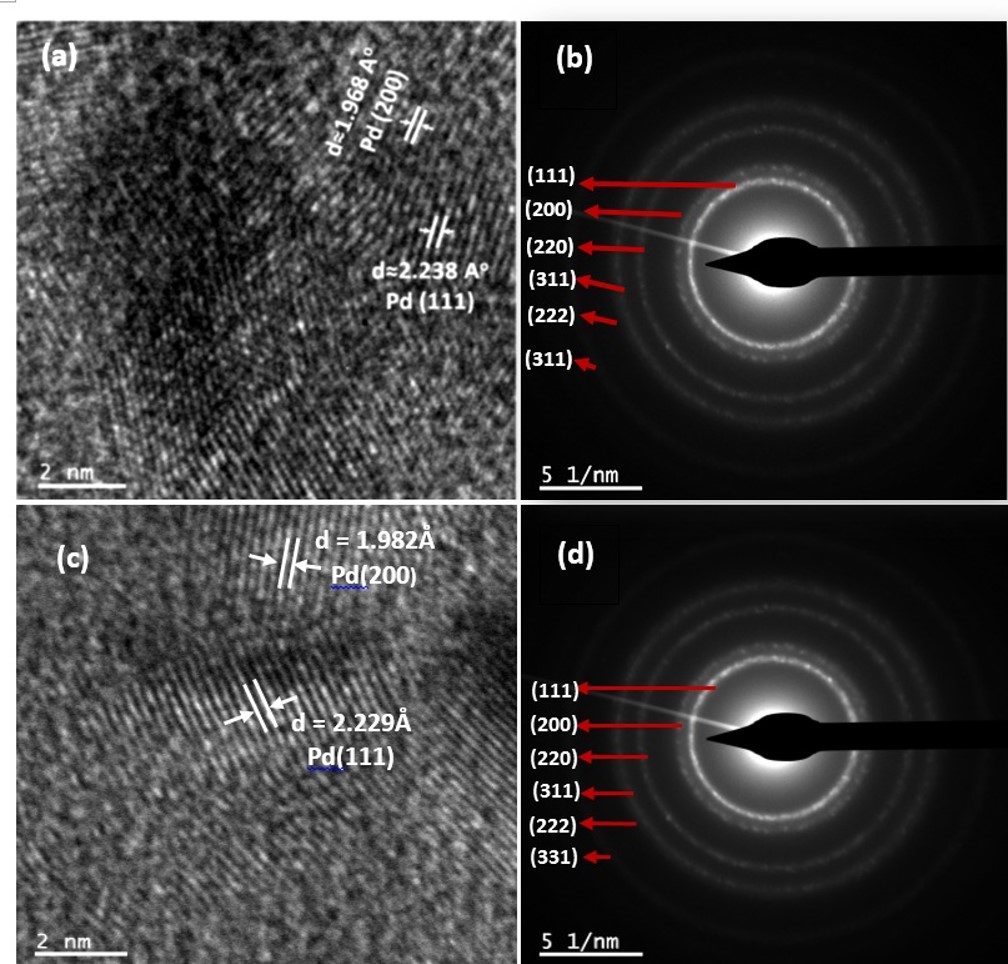}
		\caption{(a) and (c): HRTEM images of a 3 nm  and 5 nm respectively thick Pd film on a 2 nm scale. (b) and (d) represent the corresponding SAED patterns of the images (a) and (c).} 
		\label{TEM}
\end{figure}

To characterize the nano-island films for crystallinity, we measured HRTEM on each of them. Figs. \ref{TEM} (a) and (c) show representative HRTEM images measured on a 3 nm thin film and 5 nm thin film respectively, wherein, lattice planes corresponding to the formation of a uniform lattice are clearly seen. Two lattice planes, one corresponding to Pd (111) \cite{navaladian} where the distance between the lattice planes is measured to be $\approx$ 2.238 \AA ~ on an average, and the other corresponding to Pd (200) \cite{navaladian,du} where the distance between the planes is measured to be $\approx$ 1.9 \AA, have been marked. Polycrystalline nature of the films with short range ordering is evident in the ring structure of the corresponding select area electron diffraction (SAED) pattern. The ring SAED pattern are indexed to (111), (200), (220), (311), (222) and (331) lattice planes and confirm the FCC crystal structure of metallic Pd (JCPDS file No. 87-0638).

\subsection{AFM measurements}
\begin{figure}[h!]
	\centering
		\includegraphics[width=0.8\textwidth]{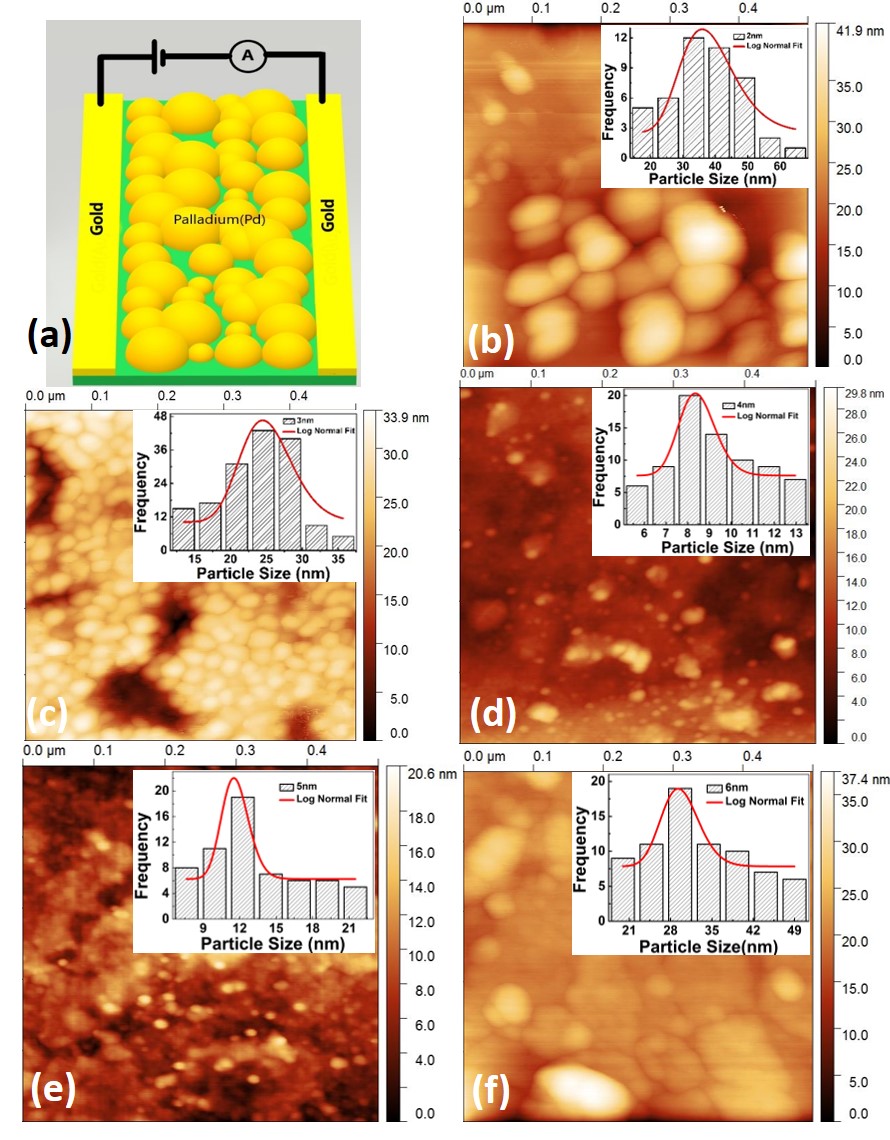}
		\caption{(Colour online): (a) Schematic of Pd nano-island thin film assembly for resistance measurements. AFM topographs obtained on (b) 2 nm (c) 3 nm (d) 4nm (e) 5 nm and (f) 6 nm thick films. Inset of each shows a log-normal distribution of nano-islands.} 
	\label{AFM}
\end{figure}

To characterize the surface morphologies of the grown films, we did AFM measurements on the grown films. Main panels of Figs. \ref{AFM} (b)-(f) show the surface morphologies of the 2 nm, 3 nm, 4 nm, 5 nm and 6 nm thick Pd films respectively. It can be seen that Pd forms a randomly connected non-uniform distribution of grains. The size-distribution of grains in each film was calculated using the Gwyddion, Image J and Scanning Probe Image Processor softwares, all of which gave consistent results. Average size of the grains in each film was obtained by fitting a log-normal distribution to the histograms obtained from the AFM images and shown as insets to the main panel of Figs. \ref{AFM} (b)-(f).

\subsection{Charge transport mechanism without hydrogen exposure}
\begin{figure}[h!]
	\centering
		\includegraphics[width=0.94\textwidth]{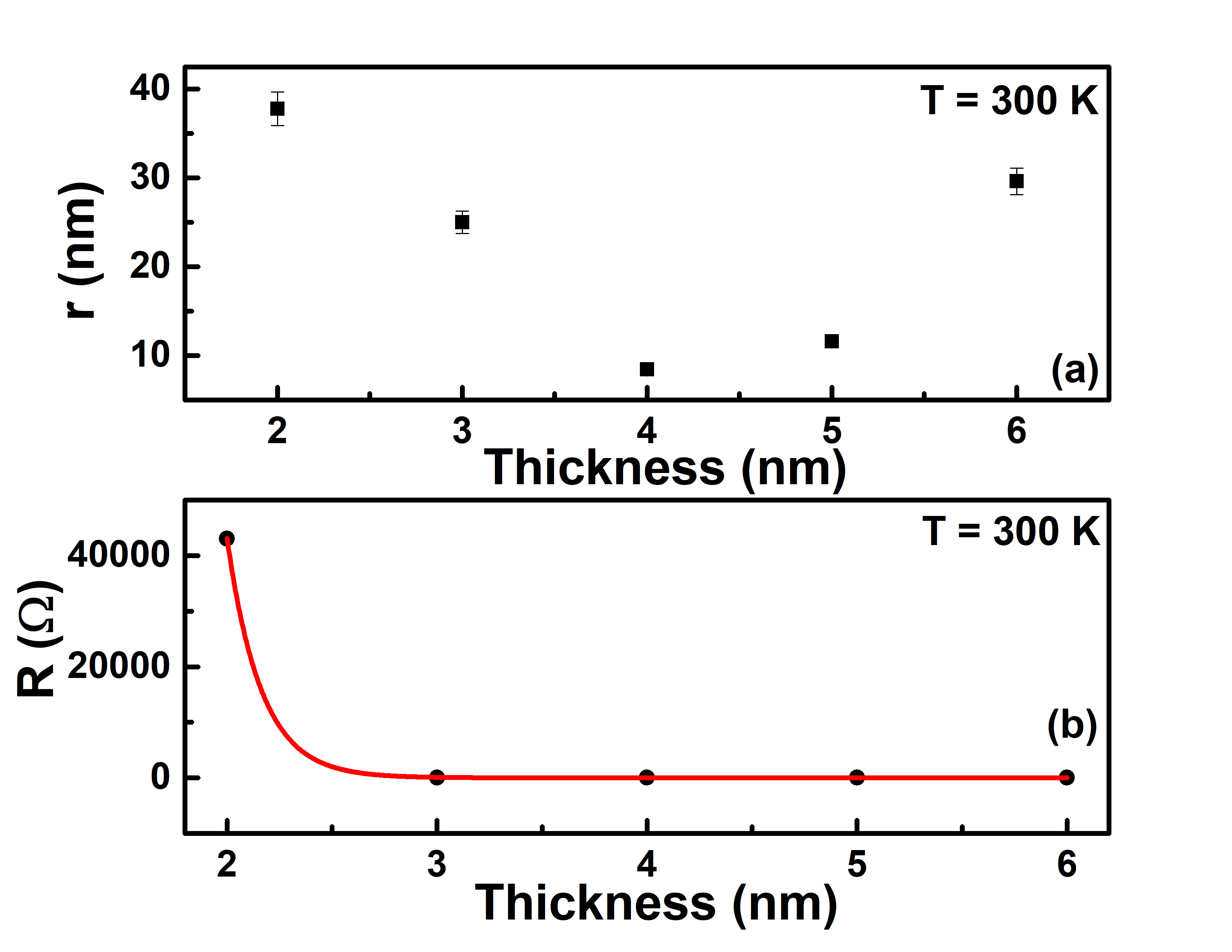}
		\caption{(Colour online): (a) Thickness variation of the calculated average size of the thin films. (b) Resistance variation of the thin films with thickness at room temperature.} 
	\label{RT-R}
\end{figure}

\begin{figure}[h!]
	\centering
		\includegraphics[width=0.8\textwidth]{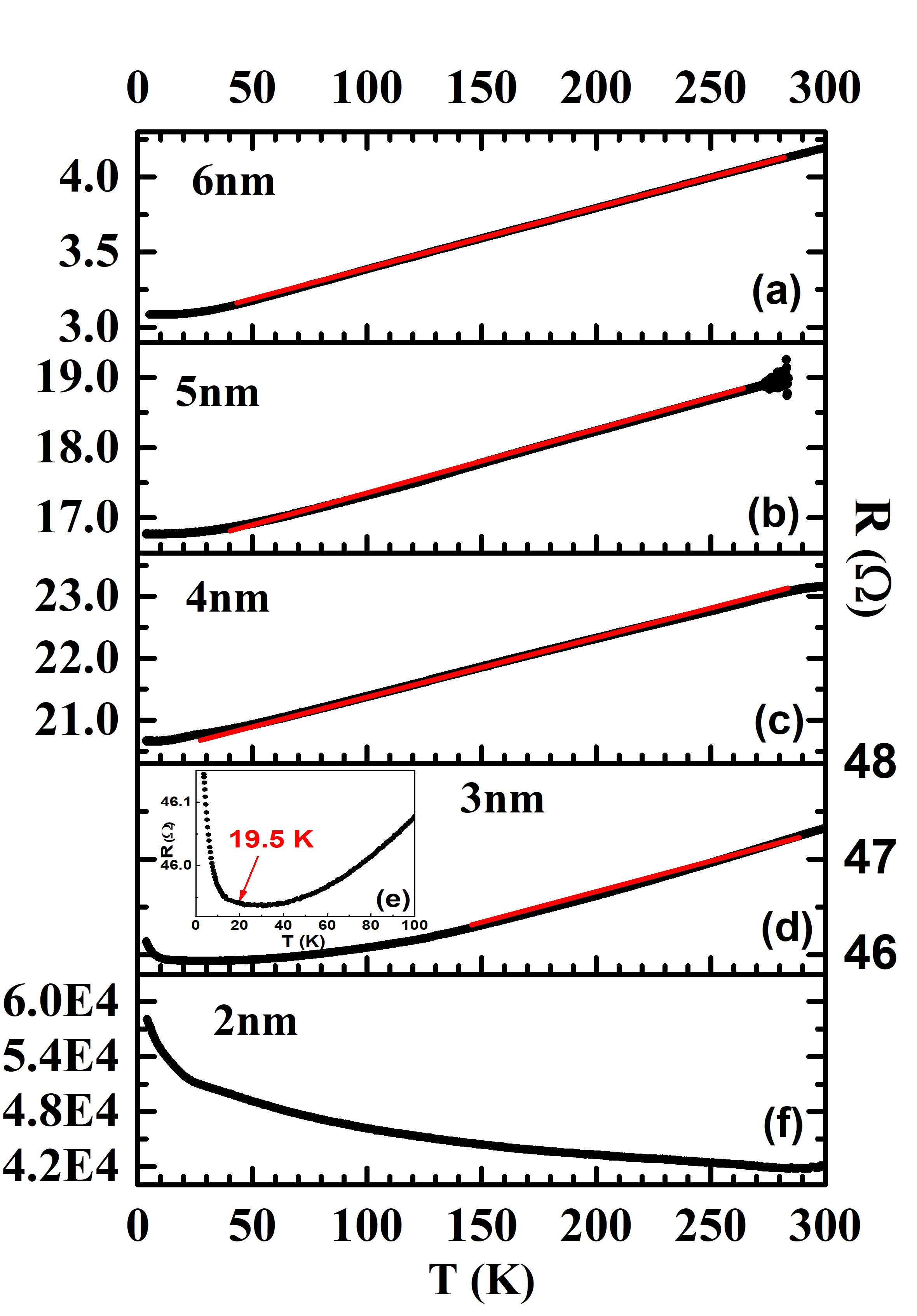}
		\caption{(Colour online): Temperature variation of resistance for the (a) 6 nm, (b) 5 nm, (c) 4 nm, (d) 3 nm and (e) 2 nm thin film. Red curves in (a)-(d) is a straight line fit. See text for details.} 
	\label{RT-d}
\end{figure}

The average particle size, \textit{r}, of the grown films obtained from the log-normal distribution curves are plotted in Fig. \ref{RT-R} (a), as a function of thickness. It can be seen that as the thickness decreases from 6 nm, \textit{r} decreases till the film thickness of 4 nm, below which, the average size starts to increase. The utilized technique of e-beam evaporation is a physical vapour deposition technique that involves condensation of a vapourised material on a substrate, wherein, various microscopic processes govern the final morphology of the grown films/nanoparticles \cite{venables}. It is known that the ratio of bulk cohesive energy, E$_c$, of the vapourised material to the adsorption energy on the substrate, E$_A$, determine if highly coalesced clusters form or thin films form. In the case of pure metal deposition, the metallic clusters are known to form in drop-like fashion, analogous to the condensation of water-vapour on a substrate \cite{beysens}. In this process, drops of metal nanoparticles keep coalescing on each other till E$_A$ wins over E$_c$ and it becomes energetically favourable for the droplets to flatten out on the substrate. The observed decrease of the average particle size with thickness from 6 nm to 4 nm, then, suggests the importance of E$_c$ in these range of thin film thicknesses while the increase in \textit{r} below 4 nm suggests the transformation of the growth process to that of surface adsorption.\\ 
In order to understand the effect of the film thickness on the electronic state of the resultant system, the room temperature resistance \textit{R} of the films are plotted as a function of thickness (d) in Fig. \ref{RT-R} (b). The black filled circles are the experimental data while the red continuous curve is a fit to the equation: 
\begin{equation} \label{eqn:RT-T}
R(d) = A*\left(\frac{1}{1-exp(-B*d)}\right)
\end{equation}

where \textit{A} is a constant and \textit{B} has a functional dependence on the thickness \textit{d}. It can be seen that the data points are fitted to the equation (\ref{eqn:RT-T}) rather well. It is well-known that the room temperature resistance of a bulk metal is dominated by scattering, primarily, due to phonons. If the thickness of the bulk metal is decreased to a thin film configuration, then other contributions are known to affect the resistance of such thin films. Primary amongst them are scattering due to grain boundaries arising due to a decrease in the size of the grains with the decrease in the thickness of the films \cite{mayadas,mayadas1}. However, it is known that if the film is very thin, then the contribution to resistance is dominated by surface scattering \cite{fuchs,sondheimer,zhang1}. In the limit of the thickness of the film reaching few nanometers, electron confinement effects are known to alter the resistance of nanostructures drastically \cite{zhang1,murray,gaponenko,mowbray,bimberg,stangl}, wherein, the metallic nature of transport may even change to that akin to insulators \cite{praveen}. The surface scattering is known to vary as (1-exp(-$\kappa$tH))$^{-1}$ \cite{fuchs,sondheimer,zhang1}, where $\kappa$ = \textit{h}/$\lambda$; $\lambda$ is the mean free path, \textit{t} is the thickness of the film and \textit{H} is a function of thickness. So, the constant \textit{B} in equation \ref{eqn:RT-T} is the product of $\kappa$ and \textit{H}. A reasonable fit of resistance to the surface scattering model, then, implies that the surface scattering dominates charge transport in these Pd thin films \cite{dutta,heiman,lacy}. It is worth noting that while the 6 nm, 5 nm, 4 nm and 3 nm thick films are metallic at room temperature, the 2 nm film exhibits insulating behavior displaying room temperature electron confinement effects therein (Fig. \ref{RT-d}). We would like to add that in order to make any quantitative estimates about the charge transport, it is necessary to use the exact resistivity expression from Mayadas and Shatzkes \cite{mayadas,mayadas1} involving terms arising due to contributions from grain boundaries and surface scattering. The solution of the expression involves numerical integration which we have not yet developed.\\
In order to understand the details of electrical transport in the thin films, temperature dependence of their resistance were measured. Fig. \ref{RT-d} (a) plots the temperature dependence of the 6 nm thick film. It can be seen that \textit{R} decreases linearly with temperature till $\sim$ 50 K below which it saturates to the residual resistance till the lowest measured temperature. This is typical of disordered metallic systems where the resistance decreases monotonically with temperature, limited by impurity or defect scattering at low temperatures \cite{zhai,mayadas,mayadas1,dutta,heiman,lacy}. A similar behavior is observed in the 5 nm thick film as well as the 4 nm thick film, as shown in Figs. \ref{RT-d} (b) and (c). According to the Matthiessen’s rule:
\begin{equation}\label{eqn:matthieseen}
R(T) = R_0 + R_1(T)
\end{equation}

where R$_0$ is the residual resistance arising due to scattering from defects and impurities in the system while R$_1$(T) corresponds to the intrinsic resistance of the metal which is temperature dependent. Several mechanisms are known to affect the temperature dependent component $R_1$, namely, elastic electron scattering, inelastic electron-phonon scattering, inelastic electron-impurity scattering, elastic electron-impurity scattering etc. giving rise to a host of interference processes \cite{gershenzon,bergmann,pritsina}. In bulk Pd metals, R$_1$(T) has been found to follow a T$^{1.5}$ behaviour from 300 K to $\sim$ 13 K \cite{white,kemp,matula} below which the temperature dependence transforms to that of T$^2$ \cite{white,schindler}.  

Black filled circles in all the graphs of Fig. \ref{RT-d} correspond to data points while red lines in Figs. \ref{RT-d} (a)-(d) is a straight line fit to equation 3 below:
\begin{equation}\label{eqn:mattheisen1}
R(T) = R_0 + \alpha T
\end{equation}
where R$_0$ is the residual resistance at 0 K and $\alpha$ is the co-efficient of the linear temperature dependence. As can be observed from the Fig. 5, a linear fit characterizes the 4-6 nm films in the higher temperature domain very well. This linear regime though progressively decreases in range below room temperature and is observed till $\sim$ 150 K in the 3 nm film. Even though the resistance decreases with temperature characterizing a metallic behavior, the observed linear temperature dependence of resistance is different from the T$^{1.5}$ behaviour that was observed in bulk metallic Pd \cite{white,kemp,matula}. However, thin films of Pd in the thickness range 15-40 nm were observed to have a linear temperature dependence in the temperature range 300 K- 100 K \cite{satrapinski}. It is to be noted that the range of linear temperature dependence is larger in thin films of 6 nm to 4 nm from 300 K to $\sim$ 50 K.  In contrast, for the 3 nm thick film, the linear variation of resistance with temperature occurs over a narrower range of 300 K to $\sim$ 150 K below which the temperature dependence of resistance becomes weaker. At $\sim$ 19.5 K, the resistance of the 3 nm film reaches a minimum, below which it starts to increase (dR/dT $<$ 0), signaling a ``metal to insulator'' transition at 19.5 K. So, even though the 3 nm film shows metallic behavior at room temperature, charge localization effects become dominant at lower temperatures ($<$ 19.5 K).
\begin{figure}[h!]
	\centering
		\includegraphics[width=0.94\textwidth]{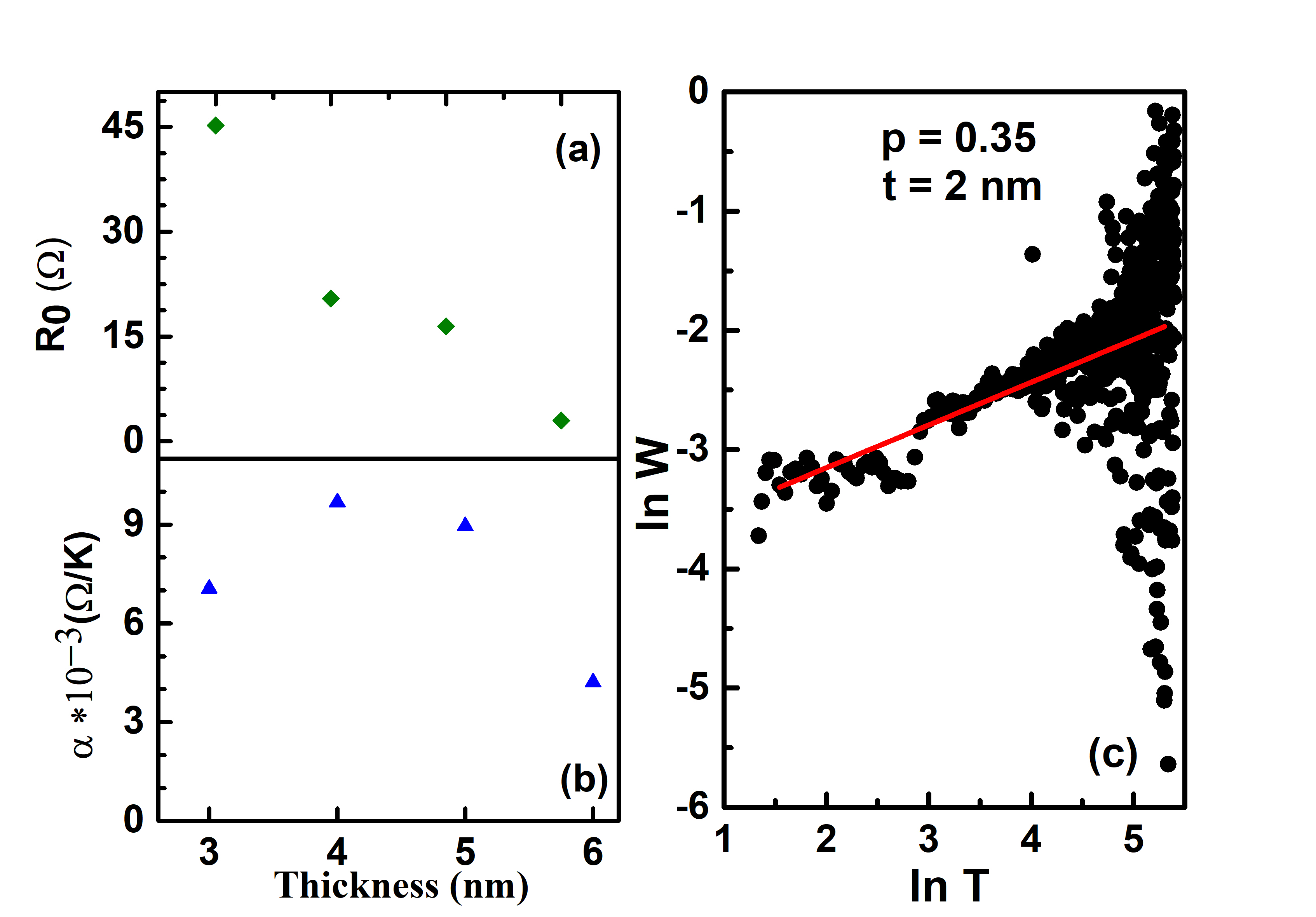}
		\caption{(Colour online): Thickness variation of (a) residual resistance R$_0$ and (b) temperature coefficient of resistance $\alpha$. (c) lnW vs. lnT plot for the 2 nm thick film. Obtained exponent p = 0.35.} 
     \label{R0}
\end{figure}

The temperature dependence of resistance of the 2 nm thick film is completely different from all the other films, in that, the resistance of the film increases with temperature in the entire temperature range of 300 K to 3.5 K (dR/dT $<$ 0), suggesting an insulator like charge transport in this film. Since the room temperature resistance is also that of an insulator ($\sim$ 43 k$\Omega$), it implies that the thin film has transformed to a discontinuous film at this thickness level of 2 nm. From Fig. \ref{RT-R} (a), we found the average particle size to be at a higher value of $\sim$ 20 nm for the film thickness of 2 nm, compared to the 6 nm thick film. This observation, then, suggests that the transformation of growth process from cohesive energy dominated for the 6 nm thin film to that of surface adsorption dominated for the 2 nm thin film happened in such a way that the inter-particles separation must have increased in the thinner film of 2 nm compared to the thicker metallic film of 6 nm, so as to give an insulating mode of charge transport for the 2 nm thick film.\\
From the fits to equation 3, the values of residual resistance R$_0$ as well as temperature co-efficient of resistance $\alpha$ were obtained for all the films. Figs. \ref{R0} (a) and (b) plot R$_0$ as well as $\alpha$ as a function of film thickness, where the symbols denote the extracted values obtained from the fit using equation 3. It can be seen that as the thickness of the film decreases from 6 nm towards 4 nm, $\alpha$ keeps increasing until the film thickness reaches 3 nm, wherein, $\alpha$ decreases. It is to be noted that for the 3 nm film, the range of straight line fit had decreased from 300 K to $\sim$ 150 K (c.f. Fig. \ref{RT-d}(d)). The non-monotonic variation of $\alpha$ with the film thickness suggests that the above mentioned processes contributing to $\alpha$, namely, elastic electron scattering, inelastic electron-phonon scattering, inelastic electron-impurity scattering, elastic electron-impurity scattering etc. \cite{gershenzon,bergmann,pritsina} together behave in a non-monotonic fashion with thickness such that one process leads the other and vice-versa as the thickness of the films is varied, such that the result is a non-monotonic variation of $\alpha$ with thickness. On the other hand, the temperature independent component R$_0$ (shown by green filled diamonds in Fig. \ref{R0} (a)) is seen to increase monotonically with a decrease in the film thickness. It is known that several factors like impurities, defects, grain boundary scattering and surface scattering contribute to R$_0$ \cite{zhai,mayadas,mayadas1,dutta,heiman,lacy}. It is expected that factors like impurities and defects that arise in a given process of making a thin film, do not vary as much as the surface whose fraction keeps increasing as the film thickness is reduced. From Fig. \ref{RT-R} (b), it was found that surface scattering contributes predominantly to the room temperature resistance as the film thickness was varied. So, the monotonic increase of R$_0$ with a decrease of film thickness in Fig. \ref{R0} (a) is ascribed predominantly to surface scattering. However, contributions from grain boundary scattering cannot be completely neglected and we ascribe the increase of R$_0$ with film thickness decrease due to a combination of predominantly surface scattering but also due to grain boundary scattering.\\ 
From Fig. \ref{RT-d} (f), it is clear that the 2 nm film is insulating at all temperatures where the resistance increased with a decrease in temperature resulting in a negative dR/dT. While studying charge transport in disordered insulators, Mott \cite{mott,mott1} found that at low temperatures, charge transport happens via electrons hopping from localized sites to localized sites near the Fermi level, such that the hopping probability is maximized at an optimal distance “r”. Assuming a constant density of states at the Fermi level, Mott found the resistance to vary with temperature as:
\begin{equation}\label{eqn:Mott}
R(T) = R_0 exp(T_0/T)^p
\end{equation}  

where
\begin{equation}\label{eqn:exponent}
p = \frac{1}{D+1}
\end{equation}  

is called the hopping exponent and \textit{D} is the space dimensionality of the solid that has a value 1/4 for a 3 dimensional solid and is equal to 1/3 for a 2-dimensional solid.\\
In order to estimate the hopping exponent, \textit{p}, it is convenient to calculate the logarithmic derivative \cite{khondaker,praveen}:
\begin{equation}\label{eqn:lnW}
W = -\frac{\partial lnR(T)}{\partial lnT} = p\left(\frac{T_0}{T}\right)^p
\end{equation} 
from where \textit{p} can be easily obtained since ln W = A-p*lnT\\
To understand if Mott’s variable range hopping is the primary mechanism of charge transport in the 2 nm thick film, we plotted lnW as a function of lnT, as shown in Fig. \ref{R0} (b). Black filled circles are the data points while the red line is a straight line fit to the data points. From the fit, the value of hopping exponent was obtained as 0.35, which is extremely close to the expected 0.33 value for a two dimensional film, thus confirming Mott’s variable range hopping as the main mechanism of charge transport in the 2 nm thick insulating Pd film.\\
From equation \ref{eqn:mattheisen1}, it is clear that the temperature co-efficient of $\alpha$ is a direct measure of dR(T)/dT, so a positive value of dR(T)/dT is an indicator of metallicity. However, a non-monotonic variation of $\alpha$, as observed in Fig. \ref{R0} (b) above, indicates that the metallicity of each film of thickness 6 nm to 3 nm is not of the pure Pd metal kind where only electron-phonon interaction is the dominant mechanism that behaves monotonically with the film thickness \cite{pritsina}. So, other mechanisms like inelastic electron-impurity scattering arising from structural disorders in the thin films of Pd, are expected to play a role in the charge transport mechasim of such films. Additionally, a metal-insulator transition at $\sim$ 19.5 K in the thin film of 3 nm thickness, is a pointer to the fact that the processes governing the charge transport in such ultra-thin films are non-trivial, do not correspond to those of bulk Pd metal, and is likely determined by structural defects/impurities, e.g., grain-size distribution etc. In order to further probe the nature of charge transport in such ultra-thin films of Pd, we decided to identify and vary alternative parameters that could have an affect on the resistance of  the films. Since Pd is known to selectively absorb $H_2$ gas, we decided to investigate the changes in the room temperature resistance of each film of Pd on its exposure to $H_2$.   

\subsection{Charge transport mechanism under hydrogen exposure}
Pd metal’s catalytic activity on H$_2$ gas and its selectivity to H$_2$ gas absorption is well known. Barr and Dankert \cite{barr,dankert} proposed a novel possibility with respect to charge transport in nano-island Pd films using the phenomenon of HAILE, wherein, the Pd islands that are close to percolation threshold swell due to H atom absorption closing the gap between islands, thereby, decreasing the resistance. This phenomenon is expected to work only for H$_2$ exposure of the order of 10,000-20,000 ppm of H$_2$ concentration since the $\alpha$ to $\beta$ transition leading to the volume expansion of the lattice is known to happen at those concentration of H$_2$ \cite{lewis,wolf,flanagan}. However, in scenarios where the inter-island gap is small enough that it could get closed during the lattice expansion of PdH$_x$ happening at the $\alpha$ $\rightarrow$ $\beta$ transition, this limitation could be overcome \cite{favier,dankert,ramanathan,xu}. Since our ultra-thin films in the 6 nm to 3 nm range are metallic (in few Ohms range) at room temperature, they are likely close to the percolation threshold and a HAILE assisted charge trasnport may be possible in such films. So, we investigated the resistive response of all the thin films, to H$_2$ gas exposure at low concentrations. For each film thickness, 3 samples were measured for a given H$_2$ concentration.
\begin{figure}[h!]
	\centering
		\includegraphics[width=0.94\textwidth]{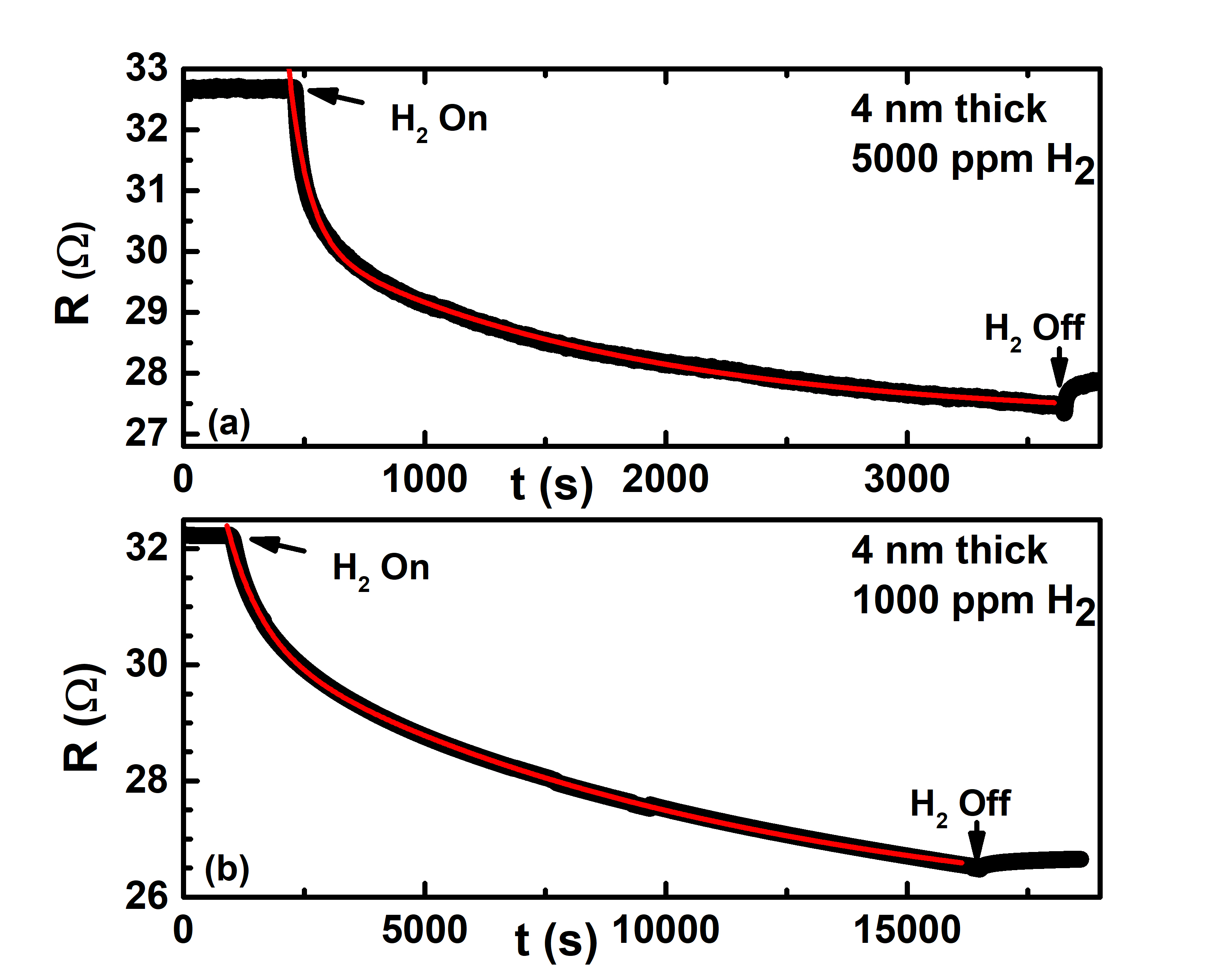}
		\caption{(Colour online): Resistance decrease in the 4 nm thick film at a H$_2$ concentration of (a) 5000 ppm and (b) 1000 ppm. Red continuous curves are fits to equation 7. See text for details.} 
	\label{R5000ppm}
\end{figure}

Figure \ref{R5000ppm} (a) shows the resistive change in a 4 nm thin film that was exposed to a  H$_2$ concentration of 5000 ppm, corresponding to 0.5$\%$ of H$_2$, shown as black solid circles. The 4 nm thin Pd film was initially exposed to an atmosphere of synthetic air in which the resistance of the film was $\sim$ 32.7 $\Omega$. As is immediately apparent from the graph, on exposure to H$_2$, the resistance decreases, likely arising from lattice expansion of Pd islands i.e. joining up of the islands. So, our nano-sized island films can detect a low concentration of 5000 ppm H$_2$, lower than the expected range of 10,000-20,000 ppm of H$_2$ concentration exploiting HAILE mechanism. By studying the effect of grain size in thin films of thickness 2 nm to 8 nm on the $\alpha$ $\rightarrow$ $\beta$ transition \cite{narehood,eastman,pundt,suleiman}, it was found that the lattice constants corresponding to PdH$_{\alpha_{max}}$ as well as PdH$_{\beta_{min}}$ increased monotonically as a function of size. Hence, the observed decrease in resistance with lower concentration (5000 ppm) H$_2$ exposure suggests that the required lattice expansion for the 4 nm thick film is of correct magnitude such that the HAILE mechanism is responsive under H$_2$ exposure.\\
In order to check if the films could detect an even lower concentration of H$_2$, we exposed the same 4 nm thick film to a H$_2$ concentration of 1000 ppm as shown in Fig. \ref{R5000ppm} (b). From the observed resistance drop from the starting value of $\sim$ 32.2 $\Omega$, it can be seen that the 4 nm thick film can detect H$_2$ even in the much lower concentration of 1000 ppm. So, our nano-island films also show the possibility of it being used as a low concentration H$_2$ gas detector. However, the films do not regain the baseline resistance value upon withdrawal of H$_2$ due to the very strong stiction of Pd on the used glass substrate. We are now in the process of overcoming this hurdle by coating the films of glass using a self-assembled monolayer that help in reducing the stiction of Pd on glass \cite{xu}.\\
Red solid lines in Fig. \ref{R5000ppm} are fits to an expression of the form:
\begin{equation}\label{eqn:tau}
R = R_a + R_b*exp(-t/\tau_1) +R_c*exp(-t/\tau_2)
\end{equation}   

where R$_a$, R$_b$ and R$_c$ are constants and $\tau_1$, $\tau_2$ are time constants defined in the usual sense denoting the time taken by the decaying resistance to fall to 1/e of its initial value. It is interesting to note that the system requires two time-constants to reach $\sim$ 83$\%$ of the starting resistance value (it takes a very long time to reach saturation due to strong stiction). The values of $\tau_1$ and $\tau_2$ obtained for a 5000 ppm exposure of H$_2$ are $\sim$ 290 s and 2780 s respectively. The values of the two time constants were found to increase to $\sim$ 500 s and 7000 s respectively for the lower concentration of 1000 ppm of H$_2$. 

\subsubsection{Percolation induced enhanced conductive pathway model}
\begin{figure}[h!]
	\centering
		\includegraphics[width=0.94\textwidth]{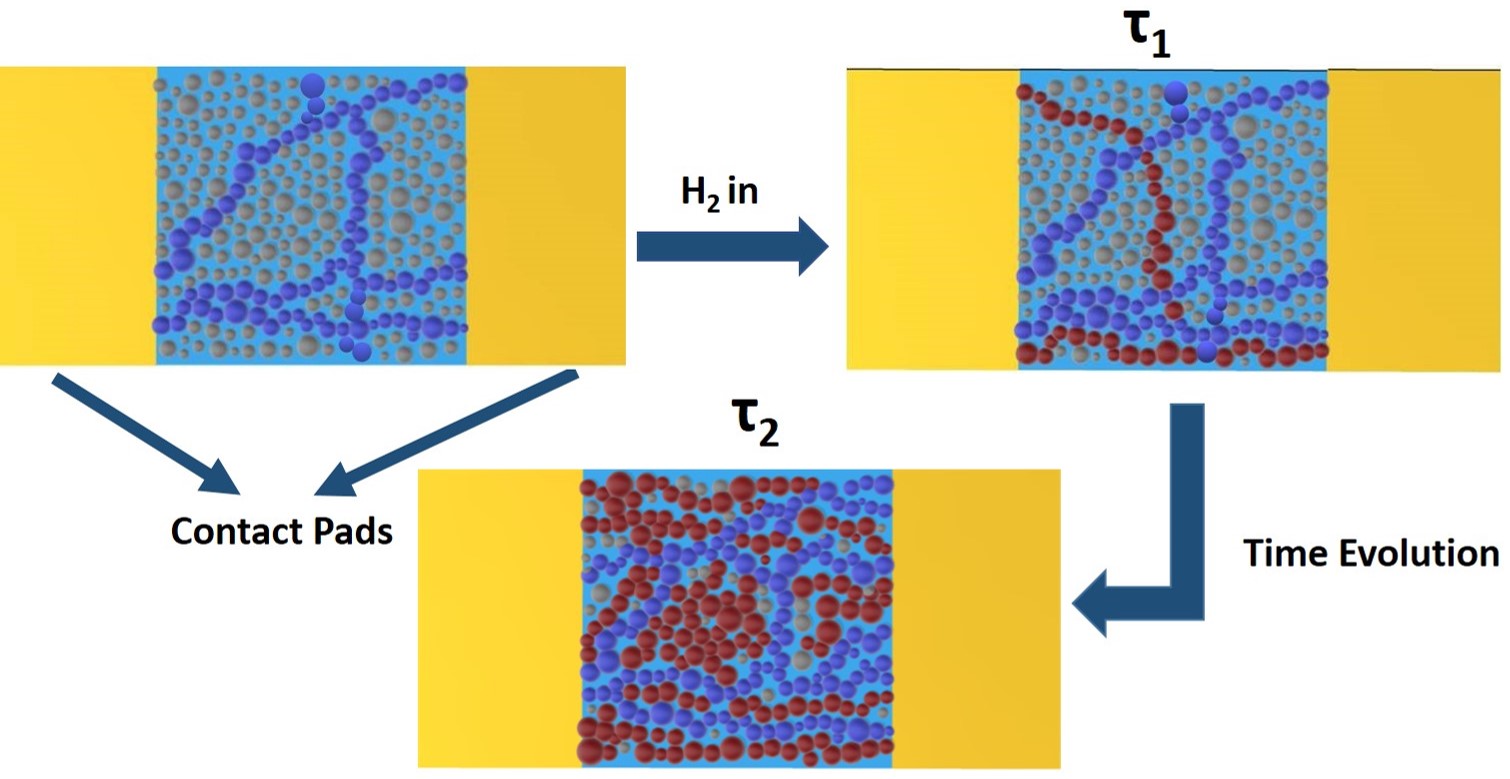}
		\caption{(Colour online): Schematic representation of enhanced conducting pathways upon H$_2$ exposure. Contact pads in each schematic is shown by yellow rectangles: (a) Before H$_2$ exposure, where grey spheres represent disconnected nano-islands of palladium and blue spheres represent at least one percolative path. (b) After H$_2$ exposure, at least one new percolative path opens at the surface resulting in the smaller time-constant $\tau_1$. (c) Opening of more percolative paths as time evolves resulting in the second time-constant $\tau_2$. } 
	\label{percolationmodel}
\end{figure}

It is known that, in general, two different time-constants in a system arise from two different mechanisms. For example, Ji et al. \cite{ji} studied the H$_2$S gas sensing property of a metal-oxide sensor SnO$_2$ and found that apart from reacting with the adsorbed oxygen anions, H$_2$S also chemisorbed onto SnO$_2$ to produce SnS$_2$. The two different mechanisms result in different time constants. Similarly, Wang et al. \cite{wang} studied gas sensing properties of Fe$_2$2O$_3$ samples with different phases, namely, $\alpha$-Fe$_2$O$_3$ and $\beta$-Fe$_2$2O$_3$, and found different gas sensing mechanisms for the two phases and, consequently, two different time-constants. However, in our case, the sensing material comprises a single element, Pd and the sensing gas, a monoatomic gas H$_2$. In such a case, the only possibility of chemisorption of H$_2$ with Pd is in making a resultant compound PdH$_x$. It is true that PdH$_x$ undergoes a series of transformation between PdH$_{\alpha}$ and PdH$_{\beta}$ as a function of the H$_2$ gas concentration. However, in the low concentration of 1000-5000 ppm of H$_2$ gas reported in this work, the resultant structure is of PdH$_{\alpha}$ type. So, different mechanisms arising from a phase change from PdH$_{\alpha}$ to PdH$_{\beta}$ doesn’t seem likely.\\    
It is known that Pd crystal’s different facets have different surface energies resulting in differences in gas sensing, and consequently, different time-constants \cite{johnson,zalineeva}. However, from HRTEM images of Fig. \ref{TEM}, we do not see the formation of any nanocrystal. So, it is also not possible that the two time-constants observed in our nanofilms may be arising from different H$_2$ loading/unloading time-scales that could have been associated with a Pd nanocrystal's different facets. So, in order to understand the presence of two time constants in the system, we propose a model based on enhanced conducting pathways in Pd nano-islands on H$_2$ exposure as shown by the schematic diagram in Fig. \ref{percolationmodel}. In hydrogen sensors made by Palladium mesowire arrays by Favier et al. \cite{favier}, two kinds of sensors were reported: “Mode I” sensors that were conductive in the absence of hydrogen and “Mode II” sensors that were insulating in the absence of hydrogen (with resistance $>$ 10 M$\Omega$). In both the kinds of sensors, namely, “Mode I” as well as “Mode II” sensors, a hydrogen exposure led to a decrease in the resistance of the sensors. In order to explain this phenomenon, the authors proposed a mechanism of hydrogen induced dilation of Palladium grains that resulted in a decrease in resistance. We have built our model based on this proposition. Our 6 nm, 5 nm, 4 nm and 3 nm thick films show conductive behavior at room temperature, while the 2 nm thick film is resistive at room temperature. So, the 6 nm - 3 nm thick films belong to the “Mode I” category while the 2 nm thick film belongs to the “Mode II” category. Grey spheres in Fig. \ref{percolationmodel} represent nano-sized palladium islands of varying sizes (c.f. Figs. \ref{AFM} and \ref{RT-R} above). Since the 6 nm - 3 nm thin films are initially conducting in the absence of hydrogen, it implies the existence of at least one percolative path for the nano-islands before any H$_2$ exposure, as shown by the connected blue spheres in Fig. \ref{percolationmodel} (a). The charge transport in such initially conductive channels may be through wavefunction overlap metallic transport. As soon as the films get exposed to H$_2$ gas, because of the HAILE process, at least one new percolative path gets formed in the nano-islands assembly (shown by the brown spheres in Fig. \ref{percolationmodel} (b)), that were initially not touching each other similar to the “Mode I” category of Favier et al. \cite{favier}. Since it is expected that the paths would be in a net-like array \cite{zeng}, so they are shown as both horizontal as well as vertical paths. Once the contact between grains is made, the charge transport may happen via wavefunction overlap metallic conduction mechanism. The opening up of the first new perlocative path reduces the resistance of the assembly and is a fast process whose time constant is governed by $\tau_1$. As time progresses, more and more new percolative paths open up, creating parallel network of resistors and decreasing the net resistance of the assembly as shown in Fig. \ref{percolationmodel} (c). The longer time constant $\tau_2$ is governed by the opening up of the many percolative paths slowly one after the other. Hence, the system takes a large time to equilibrate. This phenomenon, can also explain the extremely short time constants achieved in meso-wire arrays \cite{zeng,walter,yang} where the opening up of percolative processes is expected to happen simultaneously.\\
Since the lattice expansion from PdH$_0$ to PdH$_{\alpha_{max}}$ occurs monotonically with H$_2$ concentration \cite{narehood,eastman,pundt,suleiman}, lower concentrations of H$_2$ would require longer times for HAILE mechansim to set-up and consequently longer time constants. This is exactly what is observed in the 4 nm thick film of Pd that was exposed to a lower concentration of 1000 ppm of H$_2$ and the obtained time constants $\tau_1$ and $\tau_2$ were found to be higher than those corresponding to a higher exposure of 5000 ppm. While comparing the isotherms of a 3.8 nm and a 6 nm Pd cluster during H$_2$ loading and unloading, Pundt et al. \cite{pundt} found an enhanced solubility of H$_2$ in the low concentration region (below PdH$_{\alpha_{max}}$). Simultaneously, the lattice constants of the PdH$_{\alpha_{max}}$ structure was found to increase as compared to the bulk PdH$_{\alpha_{max}}$ such that the increase was larger in the 6 nm cluster than that in the 3.8 nm cluster. Pundt et al. \cite{pundt} attributed this observation to additional sorption of H$_2$ in the subsurface sites. As discussed earlier, surface scattering has a predominant effect on the resistance of our ultra-thin films (see discussions relevant to Fig. \ref{RT-R}). This observation, then suggests, that the enhanced detection range to lower concentrations of 1000 ppm H$_2$ observable in our ultra-thin films may be due to extra H$_2$ absorption below the surface making \textit{H} atom layers below the surface, as shown in the schematic above.  
\begin{figure}[h!]
	\centering
		\includegraphics[width=1.1\textwidth]{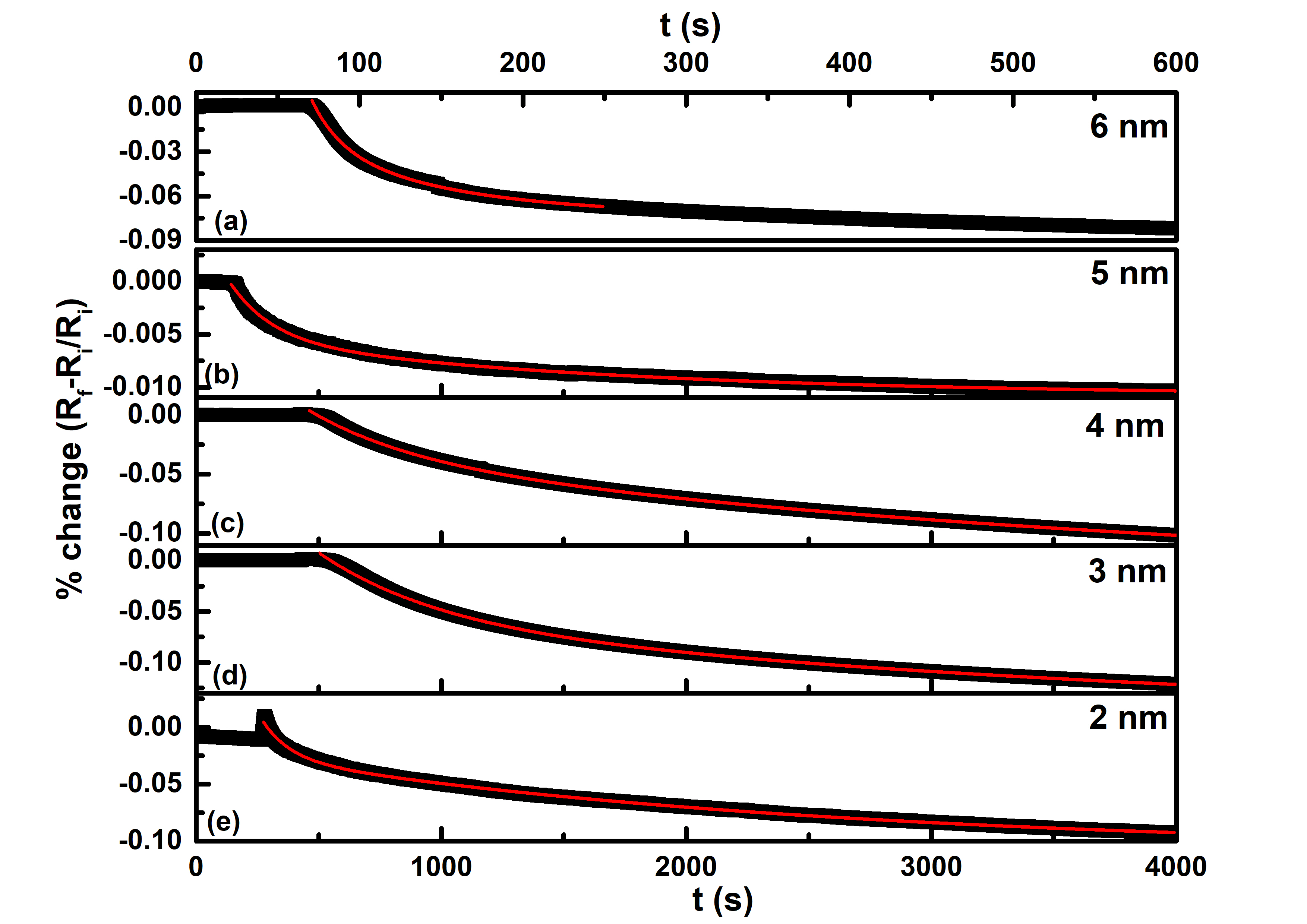}
		\caption{(Colour online): Variation of sensitivity in thin Pd films of thickness (a) 6 nm, (b) 5 nm, (c) 4 nm, (d) 3 nm and (e) 2 nm for a H$_2$ exposure of 5000 ppm concentration. Red continuous curves in each graph of (a)-(e) is a fit to estimate the time-constants $\tau_1$ and $\tau_2$.} 
	\label{tau-d}
\end{figure}

It is to be observed that the 2 nm thick film that is insulating at room temperature (c.f. Fig’s \ref{RT-R}-\ref{R0}) shows an initial increase in resistance after exposure to H$_2$ till $\sim$ 15 s after which the resistance starts to decrease again. A similar behavior was also observed by Xu et al. \cite{xu} in the 3 nm thick film that was made on the glass substrate and exposed to 2$\%$ H$_2$. They ascribed the said increase in resistance to PdH$_x$ hydride formation, since at 20,000 ppm and above H$_2$ concentrations, bulk PdH$_x$ structures transform to a lattice structure \cite{lewis,flanagan,wolf,barr,wu,morris}. However, at concentrations below 10,000 ppm, it is known that in bulk Pd, PdH$_{x}$ structures are in the form of a solid solution \cite{lewis,flanagan,wolf,barr,wu,morris}. Hence, at 5000 ppm concentration of H$_2$, the resultant PdH$_x$ structure should be amorphous, so an increase in resistance could possibly not happen due to scattering of electrons in an already disordered amorphous structure. Moreover, as already observed above, the 2 nm thick films were found to be insulating till the lowest measured temperatures and the transport in such films was found to be due to Mott’s variable range hopping mechanism. An initial increase in resistance in such films suggests that the gap between islands is too large to be closed at 5000 ppm concentration of H$_2$ exposure. The initial rise in resistance in this films, is then, ascribed to the increase in the work function of the granular Pd film due to surface adsorption of the H$_2$ molecules \cite{barr,wu,morris}, wherein, the charge gets transferred via Mott’s variable range hopping. This phenomenon is counter-acted on by the HAILE mechanism that tends to decrease the resistance of the system due to swelling up of the Pd atoms on H atom adsorption. With the passage of time, as more and more H atoms get adsorbed onto the Pd surface, Pd atoms catalytic activity decreases (due to unavailability of free Pd sites) forcing the dissociated H atoms to move from the surface inside the grain. These H atoms form additional PdH$_{x}$ structures below the surface and result in additional expansion of the Pd atoms due to swelling. At some point of time, the phenomenon of the work function increase of the surface would exactly balance the  HAILE process beyond which HAILE phenomenon would win over resulting in a subsequent decrease in resistance, as observed. For the thickest 6 nm film, since the free surface area is the least (compared to the thickness of all the other films), work-function increase due to surface would always lose over HAILE induced expansion and, hence, only a resistance decrease is observed.

\begin{table}
	\begin{tabular}{|c |c |c |}
		\hline
		\makecell{Thickness\\ (nm)}  &  \makecell{Time-constant $\tau_1$\\ (s)}  & \makecell{Time-constant $\tau_2$\\ (s)} \\
		\hline
		6   &    20.7 $\pm$ 0.08  &    --   \\ \hline
		5   &    120.2 $\pm$ 0.6  &    1081.8 $\pm$ 6.2 \\ \hline
		4   &    293.7 $\pm$ 1.2  &    2782.3 $\pm$ 25.9  \\ \hline
		3   &    340.3 $\pm$ 2  &    2883.9 $\pm$ 73.8  \\ \hline
		2   &    78.5 $\pm$ 0.2 &    1610.6 $\pm$ 2.8  \\
		\hline
	\end{tabular}
	\caption{Variation of the two time-constants $\tau_1$ and $\tau_2$ with thickness.}
	\label{Table1}
\end{table}

Our model proposes the existence of two time constants $\tau_1$ and $\tau_2$ of differing values such that $\tau_1$ has a much smaller value than $\tau_2$ since the former signals the switching on of the first new percolative path while the latter arises due to subsequent opening up of many new percolative paths. In order to see how the time-constants evolve as the thickness of the films is varied systematically from 6 nm down to 2 nm, we exposed each film to a H$_2$ concentration of 5000 ppm as shown in Figs. \ref{Table1} (a)-(e). Black curves in each graph correspond to the data points while red curves in each is a fit corresponding to equation \ref{eqn:tau}, such that $R$, $R_a$, $R_b$ and $R_c$ are replaced by their fractional change counterparts $\Delta R$, $\Delta R_a$, $\Delta R_b$ and $\Delta R_c$ respectively. The result of the fits are given in Table \textbf{\ref{Table1}}. It is to be noted that for all the other films barring the 6 nm thick film, the waiting time was for 4000 s. So, a reasonable value of the higher time-constant $\tau_2$ could be obtained. A similar value of the time-constant $\tau_2$ could not be obtained for a shorter waiting time of 600 s for the 6 nm film. From the Table \ref{Table1}, two observations can be made for the metallic films at room temperature: (i) Both the time-constants $\tau_1$ and $\tau_2$ systematically increase with a decrease in the thickness of the films and (ii) $\tau_1$ $<$ $\tau_2$ for each film such that $\tau_1$ is an order of magnitude smaller than $\tau_2$.\\
From Fig. \ref{RT-R} (a), it was found that the average size of the grains varied non-monotonically with size \textit{r}, where \textit{r} reduced with a reduction in thickness from 6 nm to 4 nm but increased below 4 nm. This variation is exactly contrasted by the behaviour of temperature coefficient of resistance $\alpha$'s variation with thickness, wherein, $\alpha$ increased with a reduction of thickness till 4 nm but decreased below it (see Fig. \ref{R0} (b)). Since various scattering mechanisms like inelastic electron-impurity scattering, elastic electron scattering etc. \cite{gershenzon,bergmann,pritsina} govern the behaviour of $\alpha$, the above observation, then, suggests that at room temperature, when the thickness is reduced from 6 nm till 4 nm, the average size \textit{r} decreases but the scattering mechanisms increases. However below 4 nm, \textit{r} increases but the scattering mechanism decrease. When hydrogen is brought in at low concentration of 5000 ppm, the resultant PdH$_{\alpha}$ structure is a disordered solid solution \cite{zuhner,bohmholdt,narehood}. It is expected that the phase formation of PdH$_{\alpha}$ would be affected by the underlying state of order of the Pd films. Consequently, the opening of percolative paths governing the time constants would strongly be dependent on this order. That the time-constants vary monotonically with a decrease in the thickness of the films indicate that the non-monotonic variation of size and $\alpha$ in completely contrasting manner result in a negation of the non-monotonic behaviour of each to produce a monotonic variation of $\tau$'s. This further implies that, even though all the films between 6 nm and 3 nm have a positive dR/dT at room temperature, the inter-grain separation of each film is increased as the thickness is decreased, requiring a longer time-constant for HAILE mechanism to set-in with decreasing film thickness. For the 3 nm thick film, the increased inter-grain distance is of such a magnitude that room temperature provides enough thermal energy for conduction. However, as the temperature is reduced, reducing the associated thermal energy, a point would be reached when this energy is insufficient for conduction, resulting in a metal-insulator transition, as observed (c.f. Fig. \ref{RT-d} (d)).\\
For the 2 nm film, in contrast, the inter-grain separation is such that even room-temperature thermal energy is insufficient for conduction and the resultant state is insulating. As mentioned above, the initial increase in resistance of the 2 nm film may, possibly, be arising due to increase in the work function of the Pd film due to surface adsorption of the incoming H$_2$ molecules at the available Pd sites \cite{barr,wu,morris,sun}. The subsequent decrease in time constants $\tau_1$ and $\tau_2$, in these films, then, suggest that the sub-surface absorption of H atoms that took place after the H atoms were pushed below the surface resulted in a state where the atoms were at a small enough distances such that HAILE mechanism set up resulted in a fast first percolation pathway formation. The observation of $\tau_1$ being an order of magnitude smaller than $\tau_2$ could be understood from the fact that $\tau_1$ corresponds to the time when the first percolative path opens which is a fast process, while $\tau_2$ corresponds to the subsequent opening up of newer percolative paths which is a slower process. 
        
\section{Conclusion:}
To conclude, charge transport mechanism of ultra-thin films of palladium of varying thicknesses from 6 nm down to 2 nm, have been investigated by studying the temperature dependence of their resistance. It was found that the films of thicknesses varying from 6 nm to 4 nm were metallic over the entire temperature range measured. The lower thickness film of 3 nm thickness was found to undergo a metal-insulator transition at a temperature of 19.5 K while the still lower thickness film of 2 nm was insulating at all measured temperatures following Mott’s variable range mechanism of charge transport. The effect of hydrogen exposure to the charge transport mechanism of the ultra-thin films was further investigated by exposing the films to low concentration ($\sim$ 5000 ppm) of hydrogen gas. It was found that at room temperature, all the metallic films exhibited a decrease of resistance on H$_2$ exposure that has been ascribed to hydrogen induced lattice expansion phenomenon. The insulating film, on the other hand, exhibited an initial increase of resistance on H$_2$ exposure that showed a decrease in resistance upon further exposure. All the films exhibited two time-constants rather than a single one to reach back to the starting value of resistance. In order to explain the presence of two time-constants in the system, we proposed a model employing enhanced conducting pathways upon H$_2$ exposure, wherein, the smaller time-constant is ascribed to the opening of the first new percolative channel and is a fast process. The higher time-constant corresponds to the slow opening up of parallel percolative channels in the ultra-thin films. So, the amount of $H_2$ gas at room temperature is as an extra parameter that could be used to tune the charge transport mechanism in ultra-thin films of Pd that may have a bearing on the use of ultra-thin films of Pd for low concentration hydrogen sensing. This study can also have implications for liquid metals which have properties that can be tuned by adding other materials into its bulk or surface \cite{kalantar,castro}. For instance, it is known that liquid metals can dissolve other elements or molecules in themselves which upon dissolution can act as precursors in the liquid environment, thereby, generating new products. Our study of the effect of dissolution of low concentration hydrogen into ultra-thin films of palladium can have manifestations on how this dissolution happens, if surfaces and sub-surface absorption is important and if opening of percolation paths similar to what has been described above have any consequence for the dissolution process.  
\section*{Conflicts of Interest} 
There are no conflicts of interest to declare.

\section*{Acknowledgement:}
D.K.S. thanks financial support from SERB, DST, Govt. of India (Grant Nos YSS/2015/001743). P.S.G. thanks ISRO RESPOND (Grant No. ISRO/RES/3/762/17-18) for financial support. D.J.-N. acknowledges financial support from SERB-DST and ISRO RESPOND, Govt. of India (Grant Nos YSS/2015/001743, ISRO/RES/3/704/16-17 and ISRO/RES/3/762/17-18). JM acknowledges financial support from SERB, Govt. of India (SR/52/CMP-0139/2012), UGC-UKIERI (184-26/2014(IC),184-16/2017(IC)) and the Royal Academy of Engineering, Newton Bhabha Fund, UK.	
     
\hfill \break

\textbf{References}\\

\end{document}